\documentclass[pdflatex,sn-mathphys-num]{sn-jnl}

\usepackage{graphicx}%
\usepackage{multirow}%
\usepackage{amsmath,amssymb,amsfonts}%
\usepackage{amsthm}%
\usepackage{mathrsfs}%
\usepackage[title]{appendix}%
\usepackage{xcolor}%
\usepackage{textcomp}%
\usepackage{manyfoot}%
\usepackage{booktabs}%
\usepackage{newunicodechar}
\newunicodechar{−}{-}
\usepackage{algorithm}%
\usepackage{algorithmicx}%
\usepackage{algpseudocode}%
\usepackage{listings}%

\theoremstyle{thmstyleone}%
\theoremstyle{thmstyletwo}%

\theoremstyle{thmstylethree}%

\begin{document}

\title[Article Title]{Optimization of photonic waveguide bends for low-index contrast material platforms }


\author[1]{\fnm{Ritu} \sur{Jangra}}\email{ee23resch12001@iith.ac.in}

\author[2]{\fnm{Kushagra} \sur{Agarwal}}\email{ee24mtech12005@iith.ac.in}

\author[3]{\fnm{Posa Harsha} \sur{Vardhan}}\email{ee22btech11214@iith.ac.in}
\author[4]{\fnm{Nandana} \sur{P K}}\email{ee24resch01001@iith.ac.in}
\author*[5]{\fnm{Naresh Kumar} \sur{Emani}}\email{naresh@ee.iith.ac.in}

\affil*[1]{\orgdiv{Department of Electrical Engineering}, \orgname{Indian Institute of Technology Hyderabad}, \orgaddress{\city{Sangareddy}, \postcode{502284}, \state{Telangana}, \country{India}}}

\affil[2]{\orgdiv{Department of Electrical Engineering}, \orgname{Indian Institute of Technology Hyderabad}, \orgaddress{\city{Sangareddy}, \postcode{502284}, \state{Telangana}, \country{India}}}

\affil[3]{\orgdiv{Department of Electrical Engineering}, \orgname{Indian Institute of Technology Hyderabad}, \orgaddress{\city{Sangareddy}, \postcode{502284}, \state{Telangana}, \country{India}}}

\affil[4]{\orgdiv{Department of Electrical Engineering}, \orgname{Indian Institute of Technology Hyderabad}, \orgaddress{\city{Sangareddy}, \postcode{502284}, \state{Telangana}, \country{India}}}
\affil[5]{\orgdiv{Department of Electrical Engineering}, \orgname{Indian Institute of Technology Hyderabad}, \orgaddress{\city{Sangareddy}, \postcode{502284}, \state{Telangana}, \country{India}}}


\abstract{Compact and low-loss waveguide bends are vital for enhancing the integration density of on-chip photonic devices. The bend loss becomes extremely critical, particularly for compact bends in low-index-contrast platforms such as Indium Phosphide (InP). To address this issue, we introduce an elliptical width-modulated bend, which exhibits low loss. For example, at 6 µm radius bend loss for elliptical width-modulated bend is 0.22 dB per 90°, resulting in ~40\% and ~27\% reduction compared to conventional circular bends and circular width-modulated bends, respectively with only a 15\% increase in footprint providing a better loss–area trade-off. Additionally, conventional effective-index-based analytical models do not account for the impact of width modulation. In this work, we develop an enhanced analytical bend-loss formulation that incorporates these effects and validates its accuracy through close agreement with full three-dimensional Finite-Difference Time-Domain (3D-FDTD) simulations. This combined analytical–numerical design strategy offers a robust pathway toward compact, low-loss, and scalable photonic circuit architectures for the next-generation of high-density integration.}

\keywords{Circular bend waveguide, elliptical bend waveguide, width modulation, photonic integrated circuits, effective index method, InP}



\maketitle

\section{Introduction}\label{sec1}

The development of photonic integrated circuits (PICs) is critical to high speed data communications \cite{bib1},  optical sensing \cite{bib2} and quantum computing \cite{bib3}. Compared to bulk optical systems, PICs offer significant improvements in size, weight, and power (SWaP) efficiency, while enabling higher performance and scalability. However, waveguide bend loss is a major obstacle to further increasing the integration density in PICs. When a waveguide has a sharp turn, part of the guided mode radiates away, causing power loss. This effect becomes significantly stronger for the small bend radii required for high-density components, including high-Q microring filters \cite{bib4}, wavelength-division multiplexing (WDM) photonic links \cite{bib5}, and optical networks-on-chip (ONoCs) \cite{bib6}. In these applications, low-loss, compact waveguides are critical for maintaining signal integrity and minimizing power consumption.

Optical power loss in bent waveguides occurs primarily due to: (i) radiation loss resulting from curvature (ii) mode mismatch between curved and straight sections, and (iii) absorption and scattering along the waveguide \cite{bib7}. Among these, radiation loss is the dominant mechanism and is strongly influenced by the refractive index contrast of the platform. Significant progress has already been made in reducing bend losses in commonly used material platforms such as silicon (Si) and silicon nitride ($\mathrm{Si_3N_4}$) waveguides, where strong optical confinement allows compact bends with minimal radiation leakage. In contrast, relatively limited research has been conducted on the loss reduction in Indium Phosphide (InP) waveguides which exhibit a lower index contrast $(\Delta = \frac{n_{\mathrm{core}}^{2} - n_{\mathrm{clad}}^{2}}{2 n_{\mathrm{core}}^{2}})$ of $\sim11\%$, compared to $\sim45\%$ in Si and $\sim20\%$ in $\mathrm{Si_3N_4}$ \cite{bib8}. As a result, InP-based PICs typically require bend radii of the order of tens of micrometers, severely limiting integration density. Despite these challenges, InP remains a preferred platform for photonic integration because it allows monolithic integration of active components (lasers, modulators, optical amplifiers) and passive components such as waveguides and splitters on a single chip in millimeter-scale chip areas \cite{bib9}. A large bend radius of waveguides increases chip footprint, limits routing flexibility and integration density. Therefore, compact and low-loss waveguide bends are essential for next-generation high-density InP integrated circuits \cite{bib10}.


Simple circular bends are often insufficient for mitigating radiation losses in integrated waveguides, as their constant curvature induces continuous leakage of the guided mode degrading the transmission efficiency \cite{bib11}. Further when the optical mode travels from the straight waveguide into the circular bend it suffers a lateral shift in the modal profile (maximum field shifts towards the outer radius), leading to the radiation loss. Euler \cite{bib12,bib13,bib14} and adiabatic bends have emerged as the new standard, effectively replacing circular bends by providing smoothly varying or non-uniform curvature that significantly improves light transmission. Early strategies to address mode mismatch between straight and curved waveguides include the adiabatic curvature-matching approach proposed by Bogaerts et al., \cite{bib15} and the introduction of lateral offsets \cite{bib16}. Although these techniques are quite effective, they often lead to large footprints that limit dense integration. Recent developments feature hybrid bends [12] and plasmonic bends \cite{bib17}, which further improve transmission by reducing radiation loss.  Bezier-curve-based bends, including recently reported modified designs \cite{bib18}, achieved lower bending loss through smooth curvature variation but their implementation requires the optimization of multiple coefficients, typically four to realize an optimal Bezier design bend. Furthermore, these methods frequently require complex redesigns for different bend radii or wavelength regimes \cite{bib19} which may limit their application in ultra-compact PICs.

Such limitations encourage the exploration of alternative bend geometries that can achieve low radiation loss at small radii without relying on complex optimization procedures or large footprints. For example, Song et al., proposed a hybrid strategy in 2020 to reduce losses in Si and $\mathrm{Si_3N_4}$ bends by combining increased waveguide width with curvature variation \cite{bib8}. More recently, Yue Zhou et al. presented a compact 90° SOI waveguide bend based on an inner circular curve and an outer Euler–circular curve, achieving a minimal loss of 0.0175 dB for an effective bend radius of 2 µm \cite{bib20}.

The geometric engineering of the bend profile presents an effective method to diminish radiation loss in integrated waveguides, in addition to traditional confinement-enhancement methods like shallow and deep etching \cite{bib21}. In this work, we address the challenge by implementing gradual width modulation in InP waveguide bends and extend this idea to elliptical bends, which require only two coefficients for optimization. The proposed modulation technique, therefore brings in an additional degree of freedom to reduce radiation loss for smaller bend radii, efficiently, enabling the design of more compact bends with improved performance and a reduced footprint. Furthermore, this approach is simple to implement and has advantages similar to other complex bend optimization techniques discussed earlier. 

This work also reports an improved analytical model for the bent slab waveguide based on Marcuse’s formulation, which now includes width modulation through an estimation of the critical radial distance along the bend.  The effective index method \cite{bib22} is integrated into this model, enabling for the quick estimation of bend loss for various waveguide configurations. The model offers both physical insight and computational efficiency, with accuracy validated through 3D-FDTD simulations, proving its utility in the design of compact, low-loss bends for dense photonic integration.

\section{Bend design and methodology}\label{sec2}

Fig.1(a) shows the three bend geometries studied in this work: the normal circular bend, the circular width-modulated bend, and the newly proposed elliptical width-modulated bend. The schematic representation of the waveguide design used for  the bend analysis is shown in Fig.1(b), where the refractive indices for the core $(\mathrm{In}_{1-x}\mathrm{Ga}_{x}\mathrm{As}_{y}\mathrm{P}_{1-y})$ and cladding (InP) are 3.58 and 3.17, respectively \cite{bib23}. 


\begin{figure}[h!]
\centering
\includegraphics[width=0.9\linewidth]{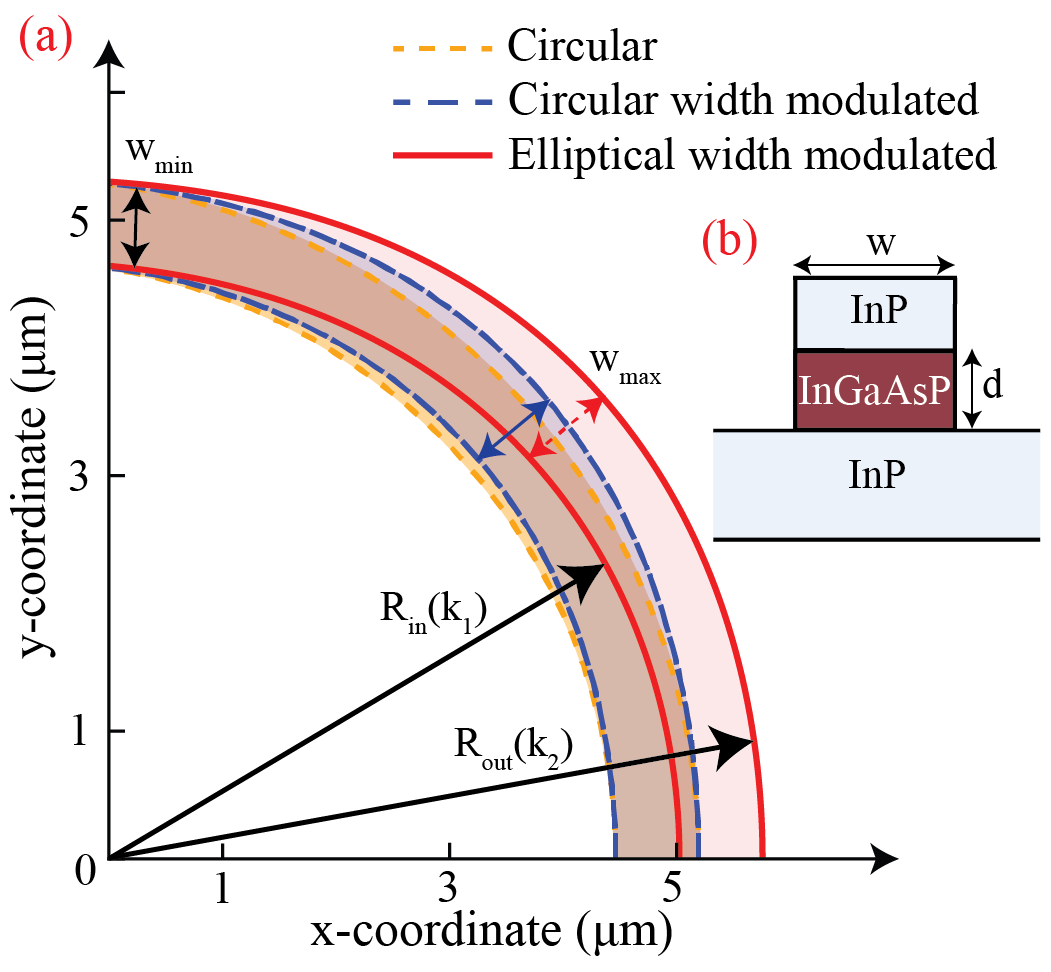}
\caption{(a) Comparison of different waveguide bend geometries: circular (dashed orange), circular width-modulated (dashed blue), and the proposed elliptical width-modulated (solid red). (b) Cross-section of the waveguide bend with an InGaAsP core of thickness $d$ and width $w$ surrounded by InP cladding.}
\label{fig:bend_geometries}
\end{figure}

\newpage
The circular width-modulated bend is designed and formulated using the approach described in \cite{bib24}. In contrast, the proposed elliptical width-modulated bend uses a continuously varying curvature to provide a smooth transition from straight to curved sections. This helps to improve the mode confinement and considerably reduces radiation leakage. The analytical representation of the elliptical width-modulated bend is expressed by the following set of equations (1-3): 
\begin{equation}
R_{\mathrm{out}}(k_1) = R_{\mathrm{ell}} + \frac{1}{2}
\left( w_{\min} + \frac{1}{k_1}\left( w_{\max} - w_{\min} \right)\eta(\theta) \right)
\end{equation}
\vspace{6pt}

\begin{equation}
R_{\mathrm{in}}(k_2) = R_{\mathrm{ell}} - \frac{1}{2}
\left( w_{\min} + \frac{1}{k_2}\left( w_{\max} - w_{\min} \right)\eta(\theta) \right)
\end{equation}
\vspace{6pt}

\begin{equation}
w_n = w_{\min} + \frac{1}{2}
\left( \frac{1}{k_1} + \frac{1}{k_2} \right)
\left( w_{\max} - w_{\min} \right)\eta(\theta)
\end{equation}
\vspace{6pt}

where $R_{\mathrm{ell}}$ refers to the distance of the midpoint of the waveguide from the origin of the bend, and it is a function of $\theta.$ The coefficients $k_1$ and $k_2$ are curvature control parameters optimized through 3D FDTD simulations to obtain the desired bend shape. In this design, the waveguide width gradually increases up to $45^\circ$ and then narrows again toward $90^\circ$. The normalized weighting function $\eta(\theta)$ maps the elliptical variation and ensures a smooth width transition, defined as:
\begin{equation}
\eta(\theta) = \frac{r(2\theta) - a_1}{b_1 - a_1}
\end{equation}

\begin{equation}
r(2\theta) = \sqrt{ \left( a_1 \cos(2\theta) \right)^2
+ \left( b_1 \sin(2\theta) \right)^2 }
\end{equation}
\vspace{6pt}

Here, $a_1$ and $b_1$ denote the semi-major and semi-minor axes with a difference of 0.001 µm, of an auxiliary reference ellipse used to define the curvature and width modulation of the bend. The function $r(2\theta)$ represents the polar radius of this reference ellipse evaluated at angular position $2\theta$ . This reference ellipse does not represent the waveguide centerline itself; instead, it is used to generate a smooth, continuously varying weighting function  $\eta(\theta)$. In this work for a 6 µm bend radius $a_1$ = 6 µm and $b_1$ = 6.001 µm, corresponding to a weakly elliptical geometry that ensures gradual curvature variation and smooth width transitions.
The minimum and maximum widths are set to $w_{\min}$ = 0.65 µm and  $w_{\max}$ = 0.85 µm, respectively, based on single-mode design constraints discussed in section 4. The optimized parameters for a bend radius of 6  µm are summarized in Table 1.

\begin{table}[htbp]
\caption{Optimized parameter values for the different bend geometries}
\label{tab:bend_parameters}
\centering
\begin{tabular}{lcccc}
\toprule
Bend type & $k_1$ & $k_2$ & $w_{\min}$ ($\mu$m) & $w_{\max}$ ($\mu$m) \\
\midrule
Circular & $\infty$ & $\infty$ & 0.65 & -- \\
Circular width-modulated & 41 & $-96$ & 0.65 & 0.85 \\
Elliptical width-modulated & 0.6 & 3 & 0.65 & 0.85 \\
\bottomrule
\end{tabular}
\end{table}

The widely used effective index method (EIM) \cite{bib25}, a standard analytical technique, is employed to model and analyze the designed InP/InGaAsP waveguides. The effective index $n_{e}$ is determined through the sequential solution of the symmetric TM and TE transcendental equations. For the bend analysis, each 90° width-modulated bend is sliced into 90 uniform angular segments, each considered as a straight waveguide of constant width, and its spatial change throughout the arc is captured by evaluating the local $n_{e}$. Among all the geometries investigated, the elliptical width-modulated bend showed the highest arc-weighted average effective index ($n_{e}$ = 3.36930), signifying enhanced optical confinement and reduced radiation loss.

\section{Analytical-Numerical bend loss analysis}\label{sec3}
In a straight channel waveguide, the fundamental optical mode is symmetrically confined within the core area. When this mode enters a curved section, it experiences a lateral shift moving away from the center of curvature shown in Fig. 2. This shift results in the radiation loss, characterized by the critical radial distance \cite{bib26}. Beyond this distance, the evanescent tail of the mode can no longer maintain phase matching with the guided field and consequently radiates into the surrounding cladding. This critical radial distance can be mathematically represented as:
\begin{equation}
x_r = R\left( \frac{\beta}{k_{clad}} - 1 \right)
\end{equation}
where R is the bend radius, $\beta$ is the propagation constant of the guided mode, and $k_2$ is the wave vector in the cladding.

\begin{figure}[htbp]
\centering
\includegraphics[width=0.9\linewidth]{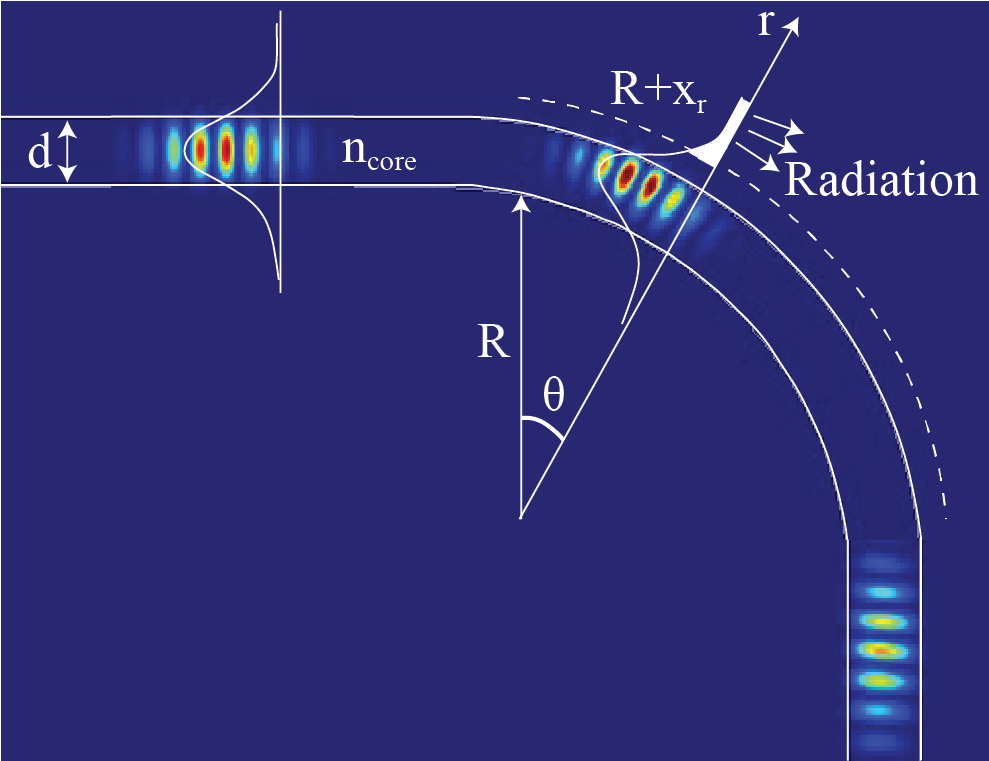}
\caption{Representation of an optical mode propagating through a bent waveguide of radius $R$. The confinement of the field inside the core ($n_{core}$) is weakened at the outer edge of the bend, causing part of the modal field to extend beyond the core and radiate into the cladding.}
\label{fig:mode_bent_waveguide}
\end{figure}

A fundamental technique to reduce bend-induced radiation loss is to increase the critical distance $x_r$, effectively shifting the radiation boundary deeper into the cladding. This reduces the fraction of optical power lost to radiation, thus increasing overall confinement. As summarized in Table 2, width modulation enhances the effective radial distance compared to conventional circular bends, where the nearly constant and minimal $x_r$ triggers early radiation onset and continuous power leakage along the bend. Among the analyzed geometries, the elliptical width- modulated bend exhibits the highest average $x_r$, confirming its superior capability to suppress radiation, particularly across the angular range of 45°-60°, where the optical path length increases more rapidly and radiation is typically pronounced. All critical distance values were calculated for each bend geometry assuming a fixed bend radius of 6 µm, and the parameters $k_1$ and $k_2$ listed in Table 1 were kept constant for all configurations.

\begin{table}[htbp]
\caption{Variation of critical radial distance and refractive index for different bend geometries}
\label{tab:critical_radius}
\centering
\begin{tabular}{lccccc}
\toprule
\multirow{2}{*}{Bend type} 
& \multicolumn{4}{c}{Critical radial distance, $x_r$ ($\mu$m)} 
& \multirow{2}{*}{$n_e$ (EIM Method)} \\
\cmidrule(lr){2-5}
& 20$^\circ$ & 45$^\circ$ & 60$^\circ$ & Avg & \\
\midrule
Circular & 0.6013 & 0.6013 & 0.6013 & 0.6013 & 3.36061 \\
Circular width-modulated & 0.6467 & 0.6921 & 0.6722 & 0.6451 & 3.36508 \\
Elliptical width-modulated & 0.7040 & 0.7324 & 0.6948 & 0.6845 & 3.36930 \\
\bottomrule
\end{tabular}
\end{table}

To complement the full 3D-FDTD simulations, the bend loss is modeled using Marcuse's analytical formulation for a curved slab waveguide \cite{bib27,bib28,bib29}. Simulation details and convergence information are provided in the Supplementary Information (Section S1) to ensure reproducibility. This model provides an exponential power attenuation coefficient $\alpha$  as a function of the bend radius R, enabling rapid estimation of bend loss while retaining strong physical insight.
\begin{equation}
\alpha_{\mathrm{bend}} = C_1 e^{-C_2 R}
\end{equation}

\begin{equation}
\alpha_{\mathrm{bend}} =
\frac{S_1 \cos^2\!\left( \dfrac{k_{1x} d}{2} \right)
\lambda_o \gamma_x^2 e^{\gamma_x d}}
{\pi n_{clad} \left( \gamma_x d + 2 \right)}
\, e^{-2 S_2 \gamma_x \left( \dfrac{\beta}{k_{clad}} - 1 \right) R}
\end{equation}
\vspace{6pt}

where $C_1$ and $C_2$ are defined as:

\vspace{6pt}

\begin{equation}
C_1 =
\frac{S_1 \cos^2\!\left( \dfrac{k_{1x} d}{2} \right)
\lambda_o \gamma_x^2 e^{\gamma_x d}}
{\pi n_{clad} \left( \gamma_x d + 2 \right)}
\qquad
\text{and}
\qquad
C_2 = 2 S_2 \gamma_x \left( \frac{\beta}{k_{clad}} - 1 \right)
\end{equation}
\vspace{6pt}

\begin{equation}
\gamma_x = k_0 \sqrt{n_e^2 - n_{eff,clad}^2}
\qquad \text{and} \qquad
k_{1x} = k_0 \sqrt{n_{eff,core}^2 - n_e^2}
\end{equation}
\vspace{6pt}

The transmission through a bend of radius of curvature $R$ and included angle $\theta$ is

\begin{equation}
\eta_{\mathrm{bend}} = \exp(-\alpha_{\mathrm{bend}} \, \theta \, R)
\end{equation}
\vspace{6pt}

and the bending loss in dB is then,
\begin{equation}
L_{\mathrm{bend}} = 4.343 \, \alpha_{\mathrm{bend}} \, \theta R
\end{equation}

\vspace{6pt}
For width-modulated cases, the bend loss is computed by numerically integrating the local attenuation along the propagation path as
\begin{equation}
L_{\mathrm{bend}} = 4.343 \int_{0}^{\pi/2} \alpha_{\mathrm{bend}}(\theta) \, R(\theta) \, d\theta
\end{equation}

Here, $\gamma_x$  denotes the decay constant in the cladding regions and $k_{1x}$ represents the propagation constant within the core. The parameters $d$, $n_e$, $n_{eff,core}$ and $n_{eff, clad}$ represent waveguide width, effective refractive index of the guided mode, refractive index of the core and the cladding regions, respectively. While the analytical model captures the dominant physical trends, its accuracy decreases for strongly confined structures or very small bend radii, where weak guidance and gradual mode deformation assumptions begin to break down. Nevertheless, the analytical predictions agree well with the FDTD simulation at an operating wavelength of 1.55 µm. The analytical curve is fitted to the FDTD data using nonlinear least squares with two fitting parameters, $S_1$ and $S_2$ as shown in equation (8), representing the bend loss variation with radius for the fundamental TE mode (Fig. 3). As expected, the normal circular bend exhibits the highest loss, while width modulation significantly suppresses radiation loss by enhancing mode confinement. This improvement arises because the waveguide width is gradually increased along the bend arc, strengthening optical confinement where curvature-induced leakage is strongest, and then smoothly tapered back to its nominal width to maintain guiding efficiency. 

\newpage

\begin{figure}[htbp]
\centering
\includegraphics[width=0.9\linewidth]{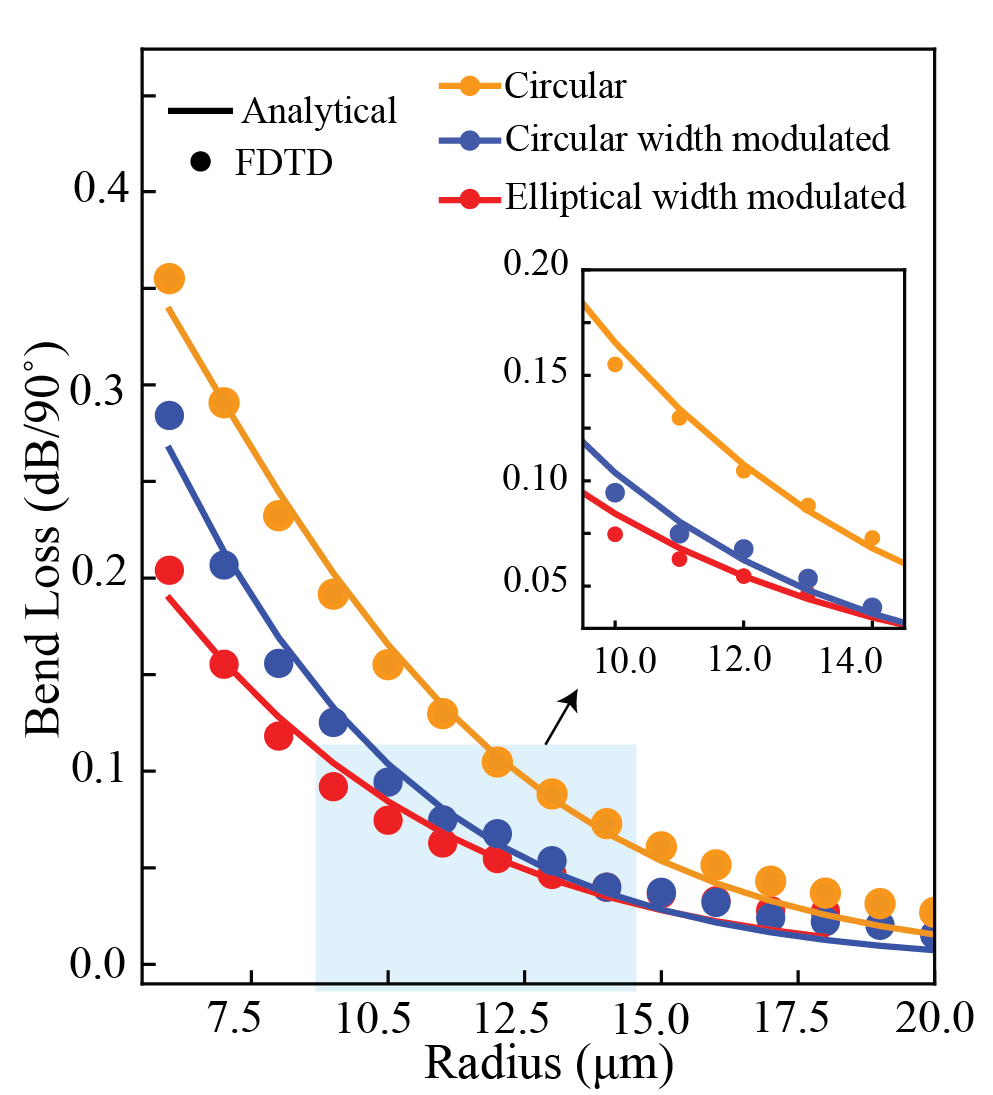}
\caption{Comparison of analytical (solid lines) and FDTD simulation (dots) results illustrating the variation of optical bend loss with bend radius for different waveguide geometries. The orange curve represents a circular bend, the blue curve corresponds to a circular width-modulated bend, and the red curve denotes the elliptical width-modulated bend. The inset highlights the low-loss region for radii ranging from 10 µm to 15 µm, demonstrating the improved performance of width-modulated designs compared with conventional circular bends.}
\label{fig:bend_loss}
\end{figure}

The gradual variation in width effectively minimizes field leakage and avoids abrupt mode distortion. With in the bend radius range of 6-13 µm, the elliptical width-modulated bend outperforms both the circular designs in performance and produces the lowest loss. The improvement is most pronounced particularly at smaller radii; for example, at R = 6 µm, the loss decreases from 0.35 dB/90° for the normal circular bend to 0.22 dB/90° for the elliptical width-modulated bend. This modulation also significantly reduces mode mismatch at the straight-to-bend transition when used in conjunction with the varying-radius (elliptical) bend profile. The effect of width modulation on the effective index profile further elucidates why elliptical width-modulated bends perform better. Unlike the normal circular bend, which maintains a nearly constant effective index along the bend, the elliptical width modulation introduces a smooth, quasi-parabolic variation in the effective index. Such graded index-like behavior facilitates the smooth adaptation of the optical mode to the lateral displacement imposed by curvature, thereby reducing the mismatch between the distributions of the inner and outer fields. Consequently, the mode suffers less abrupt distortion, the probability of radiation leakage is significantly reduced, leading to lower bending loss. Thus, the integration of geometric ellipticity in the proposed design with controlled width modulation provides an effective method for bend loss suppression, particularly under compact footprint constraints where small bend radii are required.
\begin{figure}[htbp]
\centering
\includegraphics[width=0.9\linewidth]{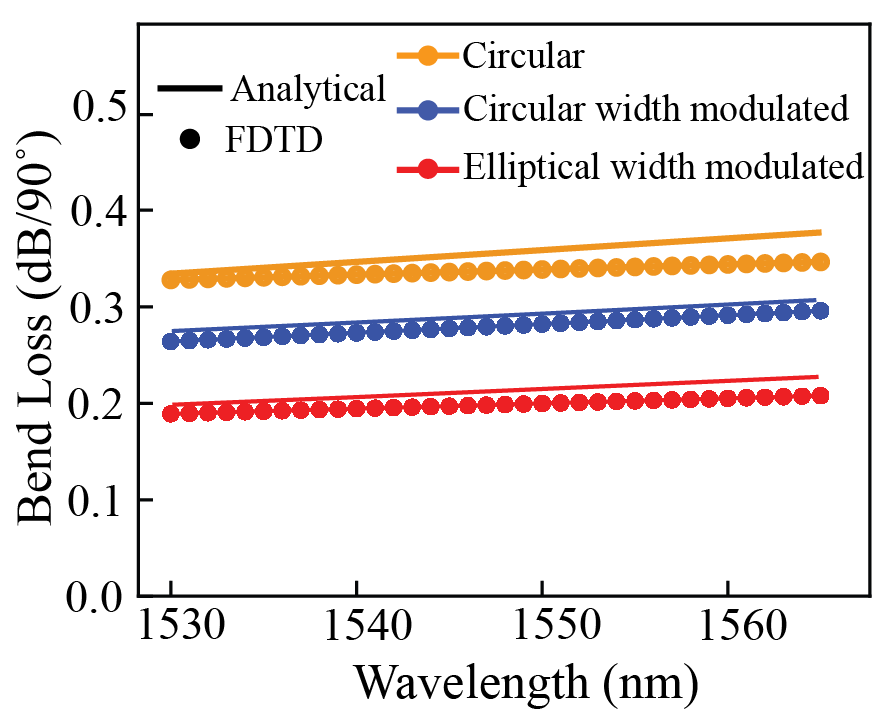}
\caption{Analytical (solid lines) and 3D-FDTD (dots) simulation comparison of bend loss versus wavelength over the C-band (1530–1565 nm) for
a 6 µm bend radius, demonstrating consistently lower loss for the elliptical width-modulated  bend compared with circular and circular width-modulated geometries.}
\label{fig:thickness_dependence}
\end{figure}
\newline
To evaluate the broadband applicability of the proposed design, the analytical model was further examined over the entire C-band ($\sim$ 1530–1565 nm). Fig. 4 shows the bend loss variation for the different geometries across the C-band for a fixed bend radius of 6 µm. The results confirm that the proposed elliptical width-modulated bend consistently suppresses radiation loss over the entire wavelength range when compared with both the circular designs. Furthermore, the analytical predictions were fitted to the corresponding 3D-FDTD results across the wavelength range, showing good agreement. This confirms that the analytical–numerical framework generalizes well over the considered spectral band. As a result, wavelength-dependent variations in modal confinement and field distribution are inherently captured by the analytical model, demonstrating that the proposed geometry retains its performance advantage over a practically relevant wavelength span. Furthermore, fabrication-tolerance analysis was performed to evaluate the robustness of the proposed bend geometries beyond the nominal design,with details provided in the Supplementary Information (Section S2).

The fitting parameters along with the associated goodness-of-fit metrics for the three bend geometries are presented in Table 3. In all cases, the high $R^2$ values ($\sim $98\%) and low root-mean-square error (RMSE) indicate that the analytical model, once properly fitted, accurately reproduces the simulation data, validating the consistency between theoretical formulation and numerical FDTD approach.

\begin{table}[htbp]
\caption{Fitting parameter values for different bend geometries}
\label{tab:fitting_parameters}
\centering
\begin{tabular}{lcccc}
\toprule
Bend type & $S_1$ & $S_2$ & $R^2$ & RMSE \\
\midrule
Circular & 0.0279 & 0.5638 & 99.07\% & 0.0094 \\
Circular width-modulated & 0.0751 & 0.6638 & 97.58\% & 0.0124 \\
Elliptical width-modulated & 0.0362 & 0.5301 & 97.56\% & 0.0088\\
\bottomrule
\end{tabular}
\end{table}

\section{Design optimization and performance metrics}\label{sec3}

The waveguide dimensions must be carefully chosen to avoid substantial bend loss while maintaining single-mode operation. Fig. 5 depicts the variation of bend loss as a function of core thickness for a bend radius of 6 µm (red, left axis). For small core thicknesses ($<$ 0.6 µm), the optical confinement is weak, leading to significant mode leakage into the cladding and increased radiation loss. As the thickness increases, bend loss decreases due to improved confinement, and reaches its  minimum near 0.85 µm. Fig. 5 also shows the normalized frequency $V$ and the normalized propagation constant $b$ (blue, right axis) where the corresponding V-number is defined as \cite{bib26}: 
\begin{equation}
V = \frac{\pi d}{\lambda}
\sqrt{n_{\mathrm{core}}^{2} - n_{\mathrm{clad}}^{2}}
\end{equation}
where $d$ is the core thickness. The normlaized propagation constant $b$ is given by,
\begin{equation}
b = \frac{n_{\mathrm{e}}^{2} - n_{\mathrm{clad}}^{2}}
{n_{\mathrm{core}}^{2} - n_{\mathrm{clad}}^{2}}
\end{equation}
indicates a shift to multimodal behaviour for core thicknesses greater than 1 µm. Hence, an optimum single-mode low-loss region between 0.65 µm and 0.85 µm is identified for the bend analyses.
For this definition of $V$, the first higher- order mode appears at $V = \pi$; therefore,the waveguide operates in single mode regime for $V < \pi$. The dashed line denotes the single-mode cutoff, while the shaded region indicates the corresponding single-mode operating range.

\begin{figure}[htbp]
\centering
\includegraphics[width=0.9\linewidth]{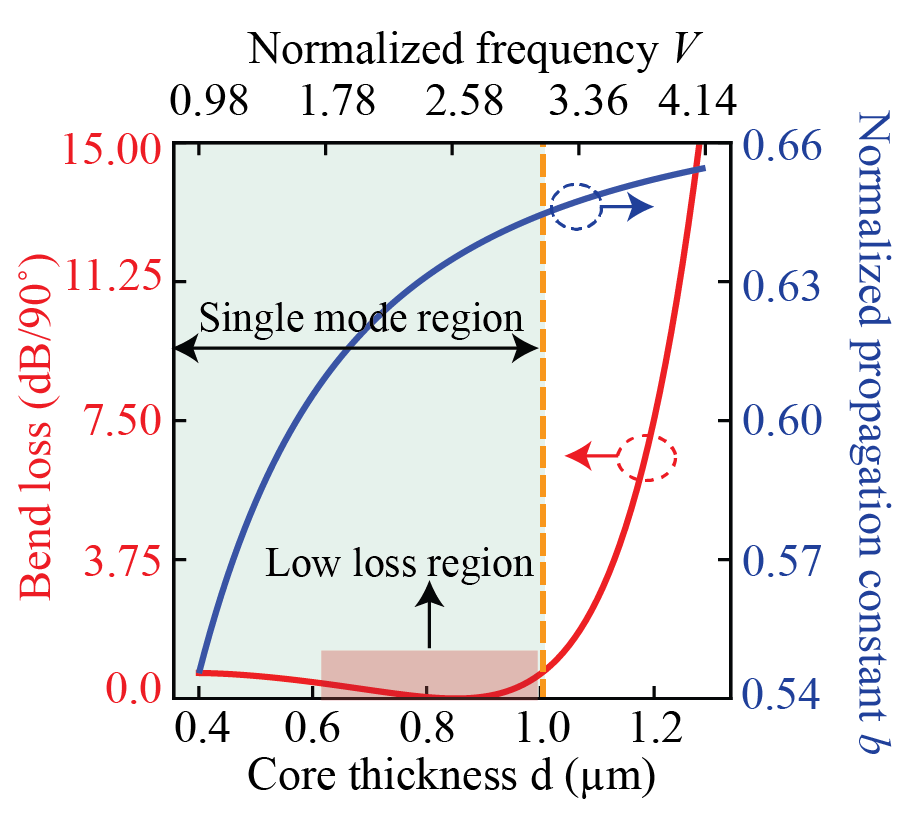}
\caption{Variation of bend loss (red curve) and normalized propagation constant (blue curve) as functions of core thickness d. The shaded regions denote the single-mode and low-loss operating regimes, while the dashed orange line indicates the single-mode cutoff condition. Optimal performance is obtained within the overlap of the single-mode (green) and low-loss regions (red).}
\label{fig:thickness_dependence}
\end{figure}

 Optimized bending shapes can significantly reduce radiation losses, they often occupy a larger physical area, which in turn limits the PIC integration density. To quantitatively evaluate this trade-off, we introduce a simple Figure of Merit (FoM) defined as loss × area, where the latter is estimated using the bounding-box method in which a rectangular area surrounding each 90° bend represents its effective layout area. Compared to the normal circular bend, the circular width-modulated reduces the loss from 0.355 dB/90° to 0.290 dB/90°, corresponding to an absolute reduction of 0.064 dB/90° (18\%). Further improvement is obtained with the  elliptical width-modulated bend, where the loss decreases to 0.22 dB/90°, yielding a total loss reduction of 0.142 dB/90° (40\%) relative to the circular bend, at the expense of an area increase of 6 $\mu\mathrm{m}^2$ (15\%) compared to both circular bend geometries. For a 6 µm bend radius, the circular bend yields a FoM of 14.2 dB·$\mu\mathrm{m}^2$, which improves to 11.6  dB·$\mu\mathrm{m}^2$ for the circular width-modulated bend and further to 10.21 dB·$\mu\mathrm{m}^2$ for the elliptical width-modulated bend. This shows that incorporating width modulation and ellipticity reduces the loss by 40 \% with only a ~15\% increase in footprint, resulting in the lowest FoM among all cases. These results confirm that the above employed strategy provides the most efficient bend geometry for dense InP photonic integration.

Direct benchmarking of the elliptical-width modulated bend with Euler, Bezier, and other adiabatic bend geometries on the InP platform is limited due to the lack of reported bend-loss data under comparable design conditions. To provide a performance context, the proposed design is compared with representative state-of-the-art bend results reported on high-index-contrast platforms with similar operating wavelengths (around 1.55 µm) and comparable minimum bend radii, as shown in Table 4. Although absolute loss values depend on material platform and confinement conditions, the reported results indicate that the proposed elliptical width-modulated bend on the InP platform achieves competitive or lower bend loss compared to reported designs on $\mathrm{Si_3N_4}$ and Si platforms for comparable radii. For small bend radii (R $<$ 13 µm), the proposed design provides substantial loss reduction, demonstrating its suitability for compact photonic integration. While the proposed design may result in a slightly larger footprint compared to certain optimized bend profiles, it provides a significantly larger reduction in radiation loss at small radii, which is critical for compact photonic integration and offers promising prospects for practical applications.

From a fabrication perspective, several practical considerations further motivate the proposed approach. For example, Bezier- and spline-based bend techniques generally require repeated numerical optimization of multiple coefficients for each change in design parameters, increasing computational complexity and design effort. In contrast, the proposed method is based on a simple analytical framework involving only two parameters, enabling rapid design adaptation without full-profile optimization. The resulting geometry is smooth, fabrication-tolerant, and compatible with standard CMOS-aligned photonic fabrication processes, facilitating practical implementation and integration.

\begin{table}[htbp]
\caption{Comparison of the proposed elliptical width-modulated bend with different state-of-the-art designs. }
\label{tab:benchmark}
\centering
\begin{tabular}{lccccc}
\toprule
Design & Platform & Waveguide dimensions ($\mu$m) & $\lambda$ (nm) & Radius ($\mu$m) & Loss (dB/90$^\circ$)  \\
\midrule

Optimal Bezier bend \cite{bib30} & Si & $0.4 \times 0.22$ & 1550 & 5 & 0.0018 (sim. )\\

Optimal NA bend  \cite{bib30} & Si & $0.4 \times 0.22$ & 1550 & 5 & 0.0004 (sim. )\\

Adiabatic bend \cite{bib31} & $\mathrm{Si_3N_4}$ & $w_{\min}=0.45$, $w_{\max}=0.7$ & 638 & 7 & 0.241 (sim.) \\

Adiabatic bend with offset \cite{bib31} & $\mathrm{Si_3N_4}$ & $w_{\min}=0.45$, $w_{\max}=0.7$ & 638 & 7 & 0.054 (sim.)\\




\textbf{This work} & InP & $w_{\min}=0.65$, $w_{\max}=0.85$ & 1550 & 7 & 0.17 (sim.)\\

Traditional Bezier bend \cite{bib32} & $\mathrm{Si_3N_4}$ & $1.2 \times 0.4$ & -- & 8 & 1.46 \\

Modified Bezier bend  \cite{bib32} & $\mathrm{Si_3N_4}$ & $1.2 \times 0.4$ & -- & 8 & 0.95 \\

\textbf{This work} & InP & $w_{\min}=0.65$, $w_{\max}=0.85$ & 1550 & 8 & 0.1424 \\

Advanced bends \cite{bib33} & $\mathrm{Si_3N_4}$ & $0.6 \times 0.3$ & 850 & 10 & 0.25 (exp.)\\

Equi-width (Euler like) \cite{bib33} & $\mathrm{Si_3N_4}$ & $0.6 \times 0.3$ & 850 & 10 & 0.33 (exp.)\\

\textbf{This work} & InP & $w_{\min}=0.65$, $w_{\max}=0.85$ & 1550 & 10 & 0.0962 \\

Inverse-designed bend \cite{bib34} & $\mathrm{Si_3N_4}$ & $0.85 \times 0.40$ & 1550 & 11 & 0.18\\ 

\textbf{This work} & InP & $w_{\min}=0.65$, $w_{\max}=0.85$ & 1550 & 11 & 0.0779 \\

\bottomrule
\end{tabular}
\end{table}

\newpage
\section{Conclusion}\label{sec3}
An analytical expression for bend loss in InP-based waveguides has been formulated by integrating the Effective Index Method (EIM) with an improved Marcuse’s radiation loss model and verifying its accuracy with 3D-FDTD simulations. The proposed model efficiently determines how bend loss varies with bend radius and curvature and allows for rapid design assessment. The proposed elliptical width-modulated bend achieved the lowest loss among the geometries studied: $\sim$0.22 dB per 90° bend at a 6 µm radius, corresponding to a $\sim 40\%$ reduction compared to a normal circular bend while only increasing the footprint by $\sim 15\%$. The analytical results are in good agreement with simulation data: ($R^2$ $\sim$ 98\%, RMSE $<$  0.0088), ensuring the accuracy of the model. Overall, this method provides a compact and physically intuitive design approach for small-footprint, low loss photonic waveguide bends in InP photonic integrated circuits that can be applied to a wide variety of photonic material platforms.

\bmhead*{Acknowledgements} This research is partially supported by the CSIR-UGC NET Fellowship and the Chanakya Doctoral Fellowship Program through the I-HUB Quantum Technology Foundation (I-HUB/DF/2022-23/06) - Government of India. It is also supported in part by the National Quantum Mission (DST/QTC/NQM/QComm/2024/2) and in part by the Core Research Grant (CRG/2022/004062) - Government of India.

\bmhead*{Author contributions} We confirm that the manuscript has been read and approved by all the named authors. Ritu Jangra (First Author): Conceptualization, Methodology, Formal Analysis, Investigation, Data Curation, Writing-Original Draft. Kushagra Agarwal: Methodology, Resources, Validation, Data Curation, Writing-Review, and Editing. Posa Harsha Vardhan: Data Curation, Formal Analysis, Review and Editing Draft. Nandana P K: Formal Analysis, Review and Editing. Naresh Kumar Emani: Conceptualization, Supervision, Funding Acquisition, Resources, Validation, Review and Editing Draft.

\section*{Declarations}

\bmhead*{Conflict of interest} The authors declare no competing interests.
\bmhead*{Data availability}  Data will be made available on request.



\end{document}